# STRING THEORY: BIG PROBLEM FOR SMALL SIZE

S. Sahoo

Department of Physics, National Institute of Technology,
Durgapur – 713209, West Bengal, India.
sukadevsahoo@yahoo.com

## Abstract

String theory is the most promising candidate theory for a unified description of all the fundamental forces exist in the nature. It provides a mathematical framework that combines quantum theory with Einstein's general theory of relativity. The typical size of a string is of the order of $10^{-33} cm$, called Planck length. But due to this extremely small size of strings, nobody has been able to detect it directly in the laboratory till today. In this article we have presented a general introduction to string theory.

**1. Introduction**

Fundamental force is a mechanism by which particles interact with each other and which cannot be explained in terms of other forces. Four types of fundamental forces exist in the nature: gravitational force, electromagnetic force, weak force and strong force [1,2]. Every observed physical phenomenon can be explained by these forces. For example, (i) gravitational force is responsible for the gravity that binds us to the surface of the earth. At a larger scale, this is responsible for binding the planets in the solar system to the sun, stars inside a galaxy and galaxies in the universe etc. (ii) Electromagnetic force acts between electrically charged particles. For example, this force holds electrons inside atoms. (iii) The weak interaction (weak nuclear force) is responsible for some phenomena at the scale of the atomic nucleus, such as beta ($\beta$) decay of radioactive nuclei in which one neutron is transformed into one proton, one electron and one anti-electron type neutrino. (iv) The strong force binds quarks together inside protons and neutrons, and also makes the latter sticks to each other to form the atomic nucleus. The salient features of these fundamental forces (or interactions) are given in Table 1.

**Table 1. Four fundamental forces**

| Fundamental Force | Carrier Particle (or Mediator) | Mass (GeV/c$^2$) | Range (m) | Relative Strength |
|---|---|---|---|---|
| Strong | Gluon | 0 | ~ $10^{-15}$ | 1 |
| Electromagnetic | Photon | 0 | ∞ | ~ $10^{-3}$ |
| Weak | $W^{\pm}, Z^0$ | 80.4, 91.2 | ~ $10^{-18}$ | ~ $10^{-13}$ |
| Gravitational | Graviton | 0 | ∞ | ~ $10^{-40}$ |

From the Big Bang model [1-4], we know that in the very early universe when temperature was very high; all the four forces were unified into a single force. Then, as the temperature dropped, the gravitational force separated first and then the other three forces separated. The time and temperature scales (1 GeV = $1.2 \times 10^{13}$ K) for this sequential separation from unification are given in Table 2. [1 GeV = $1.0 \times 10^{9}$ eV, 1 eV = $1.6 \times 10^{-19}$ Joules].

**Table 2. Sequential separation of fundamental forces**

| Characterization | Forces Unified | Time Since Big Bang | Temperature (GeV) |
|---|---|---|---|
| All 4 forces unified | Gravity, Strong, Electromagnetic and Weak | ~ 0 | ~ infinite |
| Gravity separates | Strong, Electromagnetic and Weak | ~ $10^{-43}$ s | $10^{19}$ |
| Strong force separates | Electromagnetic and Weak | ~ $10^{-35}$ s | $10^{14}$ |
| Splitting of Weak and Electromagnetic forces | None | ~ $10^{-11}$ s | $10^{2}$ |
| Present Universe | None | ~ $10^{10}$ y | $10^{-12}$ |

Physicists are trying to unify these four fundamental forces of the nature with a view to develop a fundamental theory. Till today physicists have been able to unify three fundamental forces and have constructed the *standard model* (SM) [5-9] of elementary particles. This model unifies the strong, electromagnetic and weak forces. It does not take into account the gravitational force. Due to the absence of gravity in the SM, the unification is incomplete. The gravitational force is based on Einstein's general theory of relativity. This theory is based on the principles of classical mechanics and not of quantum mechanics. Since other forces in nature obeys the rules of quantum mechanics, any theory that attempts to explain gravity as well as the other forces of nature must satisfy both gravity and quantum mechanics. Quantum mechanics works at very small distances. For example, the physics of molecules, atoms, nuclei and elementary particles can be described within the framework of quantum mechanics. But general theory of relativity describes gravity at large distances. For example, the physics of the solar system, neutron stars, black holes and the universe itself can be described within the framework of general theory of relativity.

It is seen that string theory [10-15] works very well at large distances where gravity becomes important as well as at small distances where quantum mechanics is important. This is why string theory comes into picture. Here, the statement that string theory works well at large distances means that Einstein's equations of gravity can be derived from string theory. Again at large distance means at low energy scale. In low energy limit string theory can reproduce most of the physics that we know in low energy. However, because of the presence of extra dimensions these low energy limiting theories at 4-dimension depends on the type of compactification used. String theory [16] provides a framework to address some fundamental issued in cosmology and elementary particle physics. Examples are dark matter, dark energy, supersymmetric particles, unification of all interactions, etc. For a long time string theorists have proposed an idea that there should be only one fundamental theory which describes our world and that is our string theory.

**2. String Theory**

String theory is an attempt to unify all the four fundamental forces of nature. In the standard model the elementary particles are mathematical points. But in string theory, instead of many types of elementary point-like-particles, we assume that in nature there is a single variety of one-dimensional fundamental objects known as *strings* [17]. It is not made up of anything but other things are made up of it. Like musical strings, this basic string can vibrate, and each vibrational mode can be viewed as a point-like elementary particle, just as the modes of a musical string are perceived as distinct notes. Finding the spectrum of a fundamental string is analogous to find what frequencies of sound will be produced by the musical string. The particles exist in string theory are of two types: (i) *elementary particles* which are just different vibrational modes of a single string, and (ii) *composite particles* which are made of more than one string.

Strings have characteristic length scale, which can be estimated by dimensional analysis. String theory involves the fundamental constants $c$ (velocity of light) governing laws of relativity, $\hbar$ (Planck's constant '$h$' divided by $2\pi$) related to quantum physics and $G$ (Newton's gravitational constant). This union of the three very basic aspects of the universe (i.e. the relativistic, quantum and gravitational phenomena) is embodied in three unique quantities [18]:

$$\text{Planck length } (\ell_p) = \left(\frac{\hbar G}{c^3}\right)^{1/2} = 1.6 \times 10^{-33} \, cm \text{ ,} \tag{1}$$

$$\text{Planck mass } (m_p) = \left(\frac{\hbar c}{G}\right)^{1/2} = 1.2 \times 10^{19} \, GeV/c^2 = 2.2 \times 10^{-5} \, gm \text{ ,} \tag{2}$$

and

$$\text{Planck time } (t_p) = \left(\frac{\hbar G}{c^5}\right)^{1/2} = 5.4 \times 10^{-44} \, Sec \text{ ,} \tag{3}$$

The typical length of a string is of the order of $10^{-33} \, cm$, known as Planck length. If all elementary particles are really states of a string, then why we do not see this stringy structure inside various elementary particles like electrons, quarks, etc.? The answer is that the typical size of a string is much smaller than the resolution of the most powerful microscope available today. The strongest microscope, available at the Tevatron accelerator at Fermilab, can measure the minimum distance of the order of $10^{-17} \, cm$. That is why, even if the elementary particles are states of a string, they look point-like particles.

The string theory was actually originated in the late 1960s as a theory of the strong interaction. Veneziano [19] and others derived simple formulae to describe the scattering of hadrons. Hadrons are particles interacting strongly among themselves by strong force. At that time, there were two types of descriptions available for hadron scattering [20,21]: (i) the "exchange picture" (the Regge picture) – particles can interact with others via exchange of particles, and (ii) the "resonance picture" – particles can be united during short time interval as a single particle (called a 'resonance') and then are scattered again into a system of several particles. Later it was found that both the descriptions were equivalent to each other. After the establishment of the existence of gravity in string theory, string theory was considered as a fundamental theory rather than a model for strong interaction.

A string can be either open (with two end-points) or closed (like a loop). The parameter called string tension is related to the energy density in the string theory. A string has a fixed energy density (energy/length of extension) proportional to $1/\alpha'$ and is written as $\frac{1}{2\pi\alpha' c}$, where c is the velocity of light and $\alpha'$ is the 'Regge slope' parameter [21,22]. Even when the string stretches or shrinks, the energy per unit length does not change. Hence, the total mass of a string can be determined by its total length. Classically, the length of the string in the lowest energy state is zero, and so mass of this state is also zero. Although we cannot say the length of the lowest energy states of the string is exactly zero but it is assumed that their masses are zero. It is also found that these states have non-zero spin. The massless states of closed strings contain a spin-2 state. Similarly, the massless states of open strings contain a spin-1 state. The spin-2 massless close string state behaves as graviton, which is responsible for the universal gravitational force. In the low-energy limit, it exactly coincides with the graviton. Similarly, the spin-1 open string state coincides with the gauge particles, like photon which is responsible for the electromagnetic force. Thus, string theory contains gravity and other forces.

Then question comes to our mind "How interactions take place in string theory?" Let us consider a very simple case – the electromagnetic interaction between two electrons. In this case one electron emits a photon and that is absorbed by the other electron. Since both the photon and electron are simply the same string vibrating in different modes, the emission of a photon from an electron can be considered as splitting of a string into two strings. Similarly, the absorption of a photon by an electron can be considered as joining of two strings into one. In this way, all interactions in string theory take place by splitting and joining of strings. It is observed that when end-points of two open strings touch, they join together into a longer open string. However, if the two end-points of a single open string touch, they can join together and make a closed string. Similarly, a pair of closed strings can join when any pair of points coincides and form a single closed string (Figure 1) [20].

The Planck scale is the scale that enters the Newtonian gravitational force:

$$F = \frac{1}{m_p^2} \frac{m_1 m_2}{R^2}, \tag{4}$$

which indicates

$$m_p = \frac{1}{\sqrt{G}} \text{ (in natural units } \hbar = 1, \ c = 1) \tag{5}$$

The string scale is the mass scale of string oscillator excitations:

$$m_s^2 = \frac{1}{\alpha'}, \tag{6}$$

and these are related by [23,24]

$$\frac{m_s^2}{m_p^2} = g^2 \tag{7}$$

where is g the strength of string-string interactions. The Planck scale was the natural choice for the string scale when string theory was being developed as a framework for describing strong interactions (hadron physics) [25].

**3. Different types of String Theory**

The original string theory developed in the period 1968–70 was known as *bosonic string theory*. The main disadvantages of this theory were: (i) it needs 26-dimensional space-time, (ii) the absence of fermions in the theory, and (iii) the existence of a particle with imaginary mass, called *tachyon*, generally considered as an unphysical object. In the 1970s a new type of symmetry, called *supersymmetry* was discovered. Supersymmetry [26-29] is a symmetry between fermions and bosons. Fermions are particles which have half-integral spins and obey Pauli's exclusion principle. Bosons are particles those have 0 or integral spin and do not obey Pauli's exclusion principle. Supersymmetry introduces a fermionic (bosonic) partner for every boson (fermion) differing by half a unit of spin quantum number. A supersymmetric string theory is called *superstring theory* [10-14]. There are five kinds of superstring theories: Type I, Type IIA, Type IIB, SO(32) heterotic, and $E_8 \times E_8$ heterotic. These are 10-dimensional space-time (9 spatial and 1 time dimension) theory. These five string theories differ from each other in the type of vibrations which the string performs. The salient features of the string theories are given in Table 3. These string theories have two remarkable properties [15,30]: (i) Spectrum of a string theory contain a particle which has all the properties of a graviton. (ii) These string theories do not suffer from any infinity while quantizing general theory of relativity which was one of the main problems in the point-particle theories.

Initially there were two basic problems of superstring theory: (i) these string theories can only be formulated in 10-dimensional space-time rather than the 4-dimensional space-time (3 spatial and 1 time dimension) in which we live, (ii) the existence of five different superstring theories rather than a unique consistent string theory. The first problem, having too many dimensions, is resolved via a mechanism called *compactification*. The second problem, having more than one consistent string theory, is partially resolved via a mechanism known as *duality*.

**Table 3. A brief Table of String Theories**

| Type | Space-time Dimensions | Details |
|---|---|---|
| Bosonic | 26 | Only bosons, no fermions means only forces, no matter, with both open and closed strings. a particle with imaginary mass, called the *tachyon*. |
| I | 10 | Supersymmetry between forces and matter, with both open and closed strings, no tachyon, group symmetry *SO(32)*. |
| IIA | 10 | Supersymmetry between forces and matter, with closed strings only, no tachyon, massless fermions spin both ways (*nonchiral*). |
| IIB | 10 | Supersymmetry between forces and matter, with closed strings only, no tachyon, massless fermions only spin one way (*chiral*). |
| SO(32) heterotic | 10 | Supersymmetry between forces and matter, with closed strings only, no tachyon, heterotic, meaning right moving and left moving strings differ, group symmetry is *SO(32)*. |
| $E_8 \times E_8$ Heterotic | 10 | Supersymmetry between forces and matter, with closed strings only, no tachyon, heterotic, meaning right moving and left moving strings differ, group symmetry is $E_8 \times E_8$. |

## 4. Compactification

Let us consider a simple case, a (2+1)-dimensional world in which one of the two spatial dimensions is curled up into a circle with radius R. Thus the space becomes the surface of an infinite cylinder. When R is finite, it will look like a 2-dimensional space world. If we make R very small, it will look like one-dimensional. The similar procedure will be followed in order to get 4-dimensional real world from the 10-dimensional string theory. Generally, the 9 spatial dimensions in the string theory are not all be physically observable. According to Kaluza and Klein, if we assume that 6 spatial dimensions are curled up into tiny sizes, while the remaining 3 dimensions extend to infinity (or at least to very large distances) then 9-dimensional space will be seen as 3-dimensional space. This process is known as *compactification* [15, 30-32]. It is a very complicated picture. These 6 spatial dimensions form a compact manifold, whose size is so small that we are unable to detect its existence directly.

*Calabi-Yau manifolds* are a class of 6-dimensional manifolds. When they are used as the compactified manifolds in the string theory they give rise to 4-dimensional space-time theories. Thus, compactification acts as one of the important mechanism to extract real-world physics from string theory. The detailed particle content and dynamics of the 4-dimensinal theory depends on the choice of Calabi-Yau manifolds. In fact, the Calabi-Yau manifold is responsible for the elementary particles observed in the real world. But unfortunately such extra dimensions have not been observed directly till today. These hidden dimensions are still in hiding [14].

**5. String Duality**

Initially, string theorists believed that there were five distinct superstring theories. But later on it was found that these theories are related by some kinds of transformations known as *dualities*. Duality is an equivalence relation between two string theories or two quantum field theories whose classical limits are different [31,32]. If two theories are related by a duality transformation then the first theory can be transformed in some way so that it ends up like the second theory. The two theories are then said to be dual to one another under that kind of transformation. There are different kinds of dualities exist in string theory e.g. S duality, T duality, and U duality etc.

If theory A with a large strength of interaction (strong coupling) is equivalent to theory B at weak coupling then they are said to be *S dual*. Similarly, if theory A compactified on a space of large volume is equivalent to theory B compactified on a space of small volume, then they are called *T dual*. Combining these ideas, if theory A compactified on a space of large (or small) volume is equivalent to theory B at strong (or weak) coupling, they are called *U dutal*. If theories A and B are the same, then the duality becomes a *self-duality*. For example, two of superstring theories – Type I and SO(32) heterotic – are related by S duality. Type IIB theory is self-dual. T-duality relates the two Type II [Type IIA and IIB] theories and the two heterotic [SO(32) heterotic and $E_8 \times E_8$ heterotic] theories.

**6. M – theory**

In previous section we have discussed that there are five kinds of 10-dimensional superstring theories. It is eventually seen that these five superstring theories are connected to each other and considered as different limits of a single theory. This theory is known as *M-theory* [15,30-32] (Figure 2). M-theory exists in eleven space-time dimensions. The size of the eleventh dimension is proportional to the strength of the string interactions. The dualities between the various string theories provide strong evidence that they are simply different descriptions of the same underlying theory.

Two different polynomials may be related by a change of variable e.g.

$$x^2 + 2x + 1 = t^2 - 4t + 4, \text{ for } x = 1 - t. \tag{8}$$

These two polynomials are really equivalent under a simple change of variable. Similarly, two string theories become equivalent but in mysterious ways rather than a simple change of variable.

If we compactify M-theory on a line segment we will get a 10-dimensional theory. That is, take one dimension (the 11-th dimension) to have a finite length. The end points of the line segment define boundaries with 9 spatial dimensions. We will see that the (9+1) dimensional world of the each boundary can contain strings. Although many properties of M-theory are known still now this theory is not well-understood. If string theory describes nature, then our universe corresponds to some points in the parameter space of M-theory. Finding these points is a challenging topic in particle physics in coming years.

## 7. Conclusions

String theory is an attempt to unify all the four fundamental forces by modelling all particles as various vibrational modes of an extremely small entity known as *string*. All the elements are there: quantum mechanics, bosons, fermions, gauge groups and the gravity, and in this sense be "quantum gravity". This theory requires six extra spatial dimensions beyond the real 4-dimensional world in which we live. It is assumed that these extra dimensions are curled up into so small sizes that we are unable to detect its existence directly. Till today there is no experimental confirmation of string theory. Still now there are two big problems in the string theory: (i) Typical size of the string is of the order $10^{-33}$ cm. In order to see this structure we need extremely powerful microscopes, or equivalently, extremely powerful accelerators which can accelerate particles to an energy of order $10^{19} m_p c^2$, where $m_p$ is the mass of a proton at rest. But at present we have succeeded in accelerating particles up to only an energy of about $10^3 m_p c^2$. This is why the size of strings remains beyond the dreams of experimental physics for the near future. (ii) At present, we do not even know the full equations of string theory and its complexity leads to a large number of possible solutions. But we do not know the appropriate criteria for selecting those that apply to our universe.

The fate of the string theory will be determined by its experimental predictions. We hope the Large Hadron Collider (LHC), the world's largest and highest-energy particle accelerator located at CERN near Geneva, will provide some information regarding the existence of string theory [33,34]. The LHC is expected to restart in September 2009. If the string theory is verified in experiments, it will completely revolutionize our concepts of space, time and matter. Hence, string theory is widely considered as one of the main goals of particle physics research both in theoretical and experimental in this 21$^{st}$ century.

**Acknowledgments**

We would like to thank Prof. Michael E. Peskin, SLAC National Accelerator Lab. USA; Professor Ashoke Sen, HRI, Allahabad, India, Prof. Soumitra SenGupta, IACS, Kolkata, India and Retired Prof. Narendra Nath, Kurukshetra University, Haryana, India for many illuminating discussions about string theory. We also thank Dr. Kartik Senapati, Cambridge University; Dr. Bijaya Kumar Sahoo, KVI, University of Groningen, The Netherlands; Dr. Bijaya Kumar Sahoo, NIT, Raipur and Mr. Debesh Mishra, NIT, Durgapur for their help in the preparation of the manuscript. We thank the referees for suggesting valuable improvements of the manuscript.

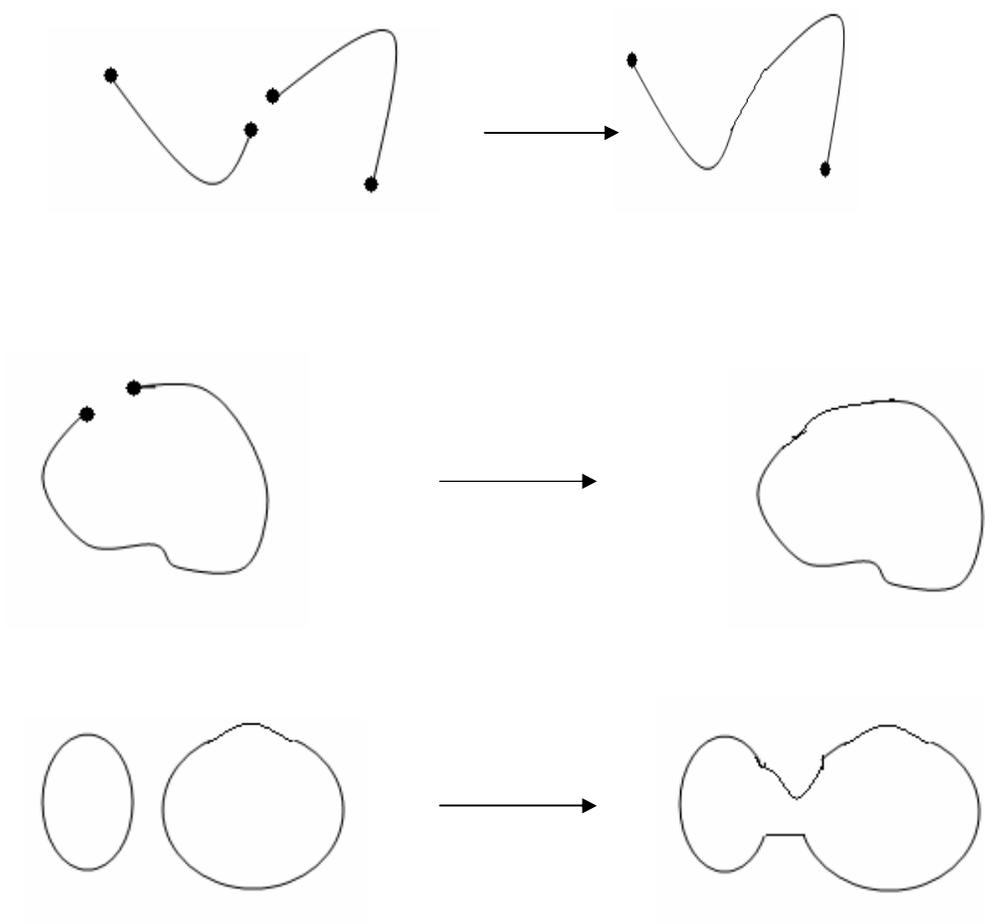

**Figure 1. Interactions of open and closed strings.**

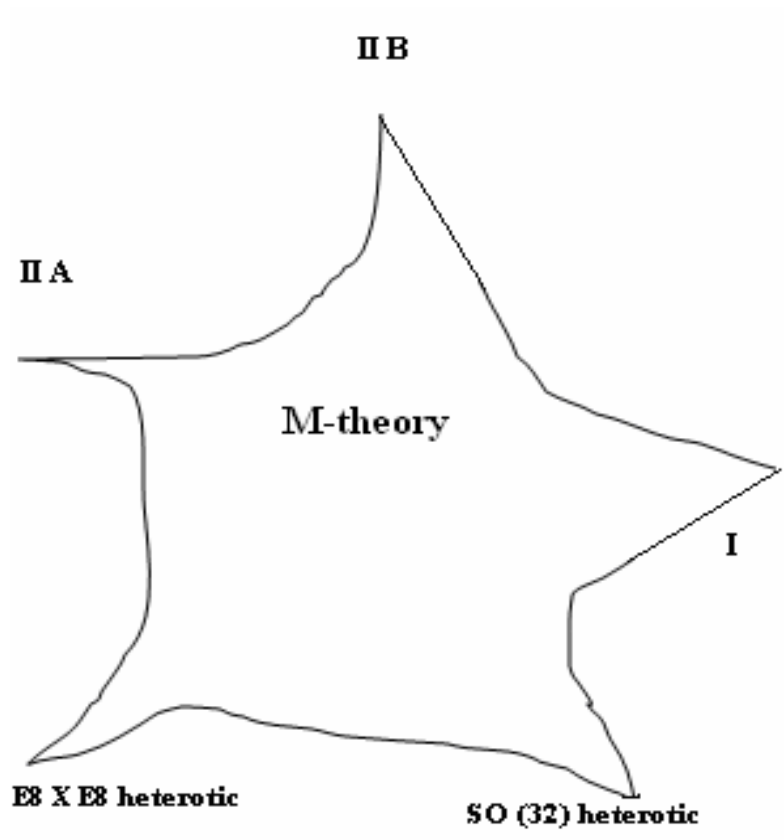

**Figure 2. Unified picture of all string theories.**